\documentclass[11pt,letterpaper]{article}

\usepackage[margin=1in]{geometry}
\usepackage[T1]{fontenc}
\usepackage{lmodern}
\usepackage{microtype}
\usepackage{setspace}
\usepackage{hyperref}
\usepackage{graphicx}
\usepackage{amsmath,amssymb}
\usepackage{cite}

\usepackage{amsmath,amssymb,amsfonts}
\usepackage[T1]{fontenc}
\usepackage{algorithmic}
\usepackage{graphicx}
\usepackage{textcomp}
\usepackage{xcolor}
\usepackage{xurl}
\usepackage{xspace}
\usepackage{listings}
\usepackage{tcolorbox}
\usepackage{subcaption}
\usepackage{caption}
\usepackage{booktabs}
\usepackage{tikz}
\usepackage{soul}
\usepackage[labelfont=bf,textfont=bf]{caption}
\usepackage{rotating}
\usepackage{enumitem}
\usepackage{multirow}
\usepackage{calc}
\usepackage{pifont}% http://ctan.org/pkg/pifont
\usepackage{array}    % for column formatting
\usepackage{adjustbox} % for tight rotation control (optional)

\def\BibTeX{{\rm B\kern-.05em{\sc i\kern-.025em b}\kern-.08em
    T\kern-.1667em\lower.7ex\hbox{E}\kern-.125emX}}

\lstdefinelanguage{Dialogue}{
    basicstyle=\ttfamily\footnotesize,
    emph = {User, Agent}, 
    emphstyle=\bfseries,
    frame=tblr,
    numbers = none,
    escapeinside={(*@}{@*)},
}

  \usepackage{hyperref}
\hypersetup{
    colorlinks = true,
    linkcolor = blue,
    anchorcolor = brown,
    citecolor = blue,
    filecolor = brown,
    urlcolor = brown
}

\usepackage{macros}

\usepackage{booktabs}
\usepackage{multirow}
\usepackage{graphicx}
\usepackage{colortbl}
\usepackage[table]{xcolor}
\usepackage{amssymb}
\usepackage{adjustbox}



\begin{document}

\title{Too Private to Tell:\\ Practical Token Theft Attacks on Apple Intelligence}

\author{Haoling (Henry) Zhou, Shixuan Zhao, Chao Wang, Zhiqiang Lin\\
The Ohio State University\\
\{zhou.3890, zhao.3289, wang.15147, lin.3021\}@osu.edu
}

\date{}

\maketitle

\begin{abstract}
\ai is a generative AI (GenAI) service provided by Apple on its devices. While offering a similar set of features as other similar GenAI services, \ai is claimed to be designed with an extra focus on user security and privacy through a two-stage authentication and authorisation design using anonymous access tokens. In this paper, we present our investigation into this token issuance mechanism with a goal to reveal possible vulnerabilities using traffic analysis, reverse engineering, and cross comparison with Apple's public documentation. 
Specifically, we present the \sysname attack, the first practical cross-device token replay attack against \ai that allows the attacker to steal the access tokens from the victim's device and utilise them on a different device, with all usage rate-limited against the victim. 
We have achieved successful attacks on the latest macOS 26 Tahoe and demonstrated that an attacker who has even used up its own allowance, can immediately regain access to \ai service. %We also performed proof-of-concept attacks on other platforms and revealed similar potential vulnerabilities. 
We have responsibly disclosed the vulnerabilities to the vendors and received confirmation from Apple with CVE assigned and bounty given. Our results highlight a general lesson for built-in AI services: Anonymising identity does not by itself make the AI service secure; Enforcing non-transferability requires cryptographic binding to the rightful user. \looseness=-1

\end{abstract}

\section{Introduction}
\label{sec:intro}

% What is Apple intelligence and Apple's focus on security and privacy
\ai is a generative AI (GenAI) service offered by Apple on their *OS devices~\cite{apple-intelligence}. When compared with other platforms, \ai has a unique focus on users' security and privacy. For GenAI requests that need to be executed on Apple's servers, Apple introduces a sophisticated workflow and an infrastructure called Private Cloud Compute (PCC) to achieve the security and privacy guarantees~\cite{pcc}. In particular, Apple significantly limits what the server can know about a particular user and device~\cite{pccdoc}, prioritising unlinkability to better protect user privacy.

% What is access tokens and how Apple's differs from others
Like most GenAI services, authentication is needed to access \ai. Common GenAI services such as OpenAI use access tokens (e.g., API keys) to verify that a request comes from an authorised user~\cite{openai-auth}. The token is also used to count usage towards a specific user for quota limitation or billing purposes. On \ai, the design is further developed in a way that can achieve the said authentication without revealing the identity of a user by issuing anonymous access tokens with a two-stage mechanism~\cite{pccdoc}. In the first stage, the service client will contact Apple's PCC Identity Service to prove that the user device is an Apple device. The PCC Identity Service will issue a long-lived Token Granting Token (TGT), which is specific to the device and the user. In the second stage, the client sends the TGT to Apple's Token Granting Service, which will issue a batch of single-use and One-Time Tokens (OTTs) to authenticate each request. Except for the PCC Identity Service, all other \ai services are proxied through third party to hide identifiable information of the user, making the two-staged authentication not traceable back to a specific user or device. \looseness=-1

% Attacks against access tokens exists and can cause big damages
Since access tokens grant access to services, they have value and are inevitably targeted by malicious attackers. In fact, token leakage for online services has been a severe issue over the years, and the reasons can range from plaintext leakage directly on open-source platforms~\cite{github-openai-key,x-api-leak} to exposure on the client side due to insecure designs~\cite{appsecret,api-leak}. If a malicious attacker is able to get the access token, it is capable of pretending to be the victim and use the service in its own interest but have all usages counted towards the victim's bill. The cost of such leak can be high~\cite{api-leak-cost} and could be even more severe on modern GenAIs.

% Attacks against Apple Intelligence is even more serious because...
Due to the nature of \ai services and how widely available they are on Apple devices, the impact of an attack against \ai's access tokens and the consequences are significantly more serious than previous attacks targeting other platforms: First, the victim, as the owner of the access tokens, is a regular consumer with no professional knowledge in cybersecurity, while previous attacks targeted keys owned by apps developed by professional developers. % Go deeper
Second, \ai's access tokens are essentially bearer tokens, which means whoever possesses them will be granted access to services. %\sx{What do you mean by bearer tokens?} \hz{Bearer token is the type of access token in authentication that grants access to whoever posses it. Unlike access tokens that are cryptographically signed, I mean this nature of the Apple's authentication token could be a weakness.}\sx{You need to introduce the terms first unless they are obvious} 
Along with the no login feature, there is no mechanism to distinguish an attacker who owns the stolen bearer token from a regular user with a legitimate device and rightfully issued token unless the authentication requires device information. %\sx{We know it requires device info so you just don't talk about the no login. There's a hidden d facto login using device info.} \hz{How no login becomes a d facto? All other GenAI services, such as Copilot, ChatGPT, requires login. Apple is the first so isn't it worth mentioning?} 
Third, there is no proper way for a victim to revoke the access tokens, meaning that once the tokens are stolen, the victim will need to wait until the next scheduled token refresh, while in previous attacks the victim could revoke an access token at any time. If such an attack exists and is easy enough for large scale deployment, an attacker would be able to \textit{harvest} these tokens to form a pool of endless GenAI resources, all under cover of the privacy and security offered by Apple itself. Naturally, we ask the question:

\begin{quote}
    \textit{Is it possible for an attacker to easily steal the tokens from a victim's Mac and use them on another Mac to disguise it as the victim?}
\end{quote}

% We found an attack
Apple has been well-known for their security protections on their devices~\cite{ios-security} and a user would certainly have expected the same level of protection for these access tokens. However, we have to unfortunately give a yes to the above question. In this paper, we present \sysname, an attack that breaks the intended non-transferability of Apple Intelligence credentials. 
%In our threat model, the victim downloads and runs an attacker-distributed application that requests keychain access, which requires the user to click ``Allow’’ on a standard macOS authentication prompt. Under this model, 
A malicious adversary can extract bearer access tokens from a victim’s Mac and replay them on another machine, thereby gaining service access and spending quota under the victim’s entitlement. We have implemented the attack and demonstrated the feasibility on the latest macOS 26 Tahoe. We demonstrated that an attacker who has used up its own request allowance can immediately regain access to \ai services by importing the access tokens stolen from a victim Mac. We have ethically disclosed this vulnerability to Apple with details of the attack. 

In summary, we make the following contributions:
\begin{packeditemize}
    \item \textbf{Systematic Understanding.} We performed a systematic anatomy of the authentication flow of \ai platform with traffic decryption and analysis, reverse engineering, and cross comparison with public documentation from Apple. Our findings pinpointed the critical steps in the authentication flow and led to the discovery of its vulnerability.\looseness=-1
    \item \textbf{Practical Attack.} We proposed \sysname, the first attack against \ai's authentication process with minimal end-user interactions that has the potential to be deployed at scale. We demonstrated that the attacker can use a victim's tokens at will even if the attacker's own allowance has been used up. %We also investigated and showed the potential of applying \sysname to other platforms.\looseness=-1
    \item \textbf{Countermeasures.} In addition to disclosing the attack to Apple and other vendors, receiving confirmation as well as a bug bounty with CVE assigned from Apple, we also discussed why \sysname would work to offer insights into potential mitigations and techniques that can be applied to similar scenarios. The idea of Apple's fix in its macOS 26.2 security patch was closely related to one of our proposed approaches. %\sx{macOS 26.2. Also Apple's approach is not exactly our proposed technique. It may have been inspired by our proposed approaches but not exactly the same. Don't overclaim.} \hz{I think you are being too defensive. In \autoref{subsec:depriv} we clearly wrote `Mitigation Adopted by Apple'. People don't bother verify this claim against Apple. Also it is true that Apple changed their mind after you gave them proposed fix. I don't see this as an overclaim.} \sx{DO NOT MAKE FACT ERROR. IF SOMETHING IS NOT TRUE THEN DO NOT CLAIM IT. We don't need the credit of fixing it. If we say Apple fixed the stuff using a method close to the concept we proposed then that's fact. If we say Apple fixed the stuff using our proposed technique then we're lying.}
\end{packeditemize}

\section{Background}
\label{sec:background}

\subsection{\ai and PCC}

\ai is a GenAI service offered by Apple on macOS, iOS, iPadOS, and visionOS. It is an integrated component of the OS itself on the client side, and can run autonomously for tasks like summarising the notifications or be invoked by the user for tasks like generating pictures~\cite{apple-intelligence}. While several lightweight tasks can be done using the on-device models, complex requests require cloud services provided by Apple to offload the heavy computation.

\begin{figure}
    \centering
    \includegraphics[width=.6\linewidth]{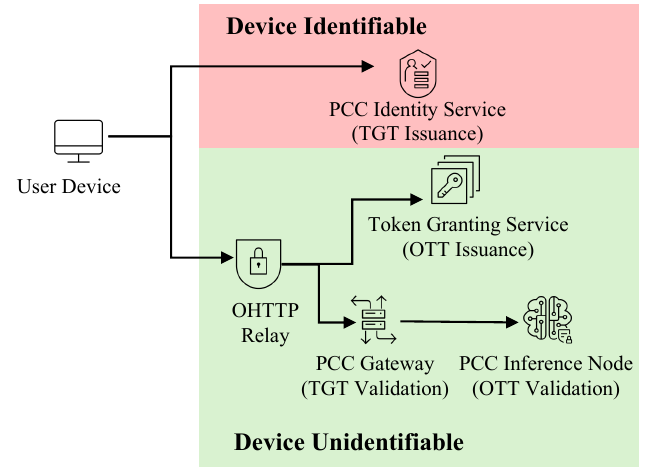}
    \caption{Security components of \ai}
    \label{fig:ai-sec-comp}
\end{figure}

% Just jump into PCC is more straightforward. No need to talk about Apple's reason
% One of the key features of \ai's marketing is about security and privacy, particularly for requests that are handled by Apple's server in the cloud~\cite{pcc}. 
Different from conventional GenAI services, Apple introduced Private Cloud Compute (PCC), a design that combines a privacy-preserving workflow that hides a user's identifiable information and an iOS-like inference environment to guard the execution against attackers. We illustrate key security measures of PCC documented by Apple~\cite{pccdoc} in \autoref{fig:ai-sec-comp} which include:
\begin{packeditemize}
    \item \textbf{OHTTP Relay.} Apple's PCC employs a variant of Oblivious HTTP (OHTTP) protocol~\cite{ohttp} in a similar fashion as their iCloud Private Relay~\cite{private-relay}, which is serviced by third parties including Cloudflare and Fastly. Since all the requests are sent via this relay, \ai services cannot tell identifiable metadata like the user's IP addresses. All user interactions with \ai services go through this relay except for the PCC Identity Service, which is used to verify if the user is using an Apple device eligible for \ai.
    \item \textbf{Access Tokens.} In the design of PCC, Apple uses a two-stage token issuing mechanism to achieve a privacy-preserving authentication workflow, following the protocol RFC 9578 Privacy Pass~\cite{privacy-pass}. Apple separates the token issuing by an anonymous but per-device Token Granting Token (TGT) and a single-use, totally anonymous One-Time Token (OTT). This detaches the device eligibility verification from the actual usage authentication and therefore makes each request unlinkable from the actual user. Since this flow is highly related to \sysname, we will introduce more details in \autoref{sec:workflow}.\looseness=-1
    \item \textbf{PCC Nodes.} A user's request is handled by a special kind of server called a PCC node. A PCC node is essentially a Mac running a customised iOS but has minimised components installed and a strong restriction on software executed on it. Hardware wise, a PCC node is manufactured with supply chain attack prevention and has a tamper switch activated once it is sealed in the factory. Software wise, it uses security features similar to those found in iPhones~\cite{pccdoc}: SPTM, TXM, AMFI, etc., which are well studied in previous literatures such as~\cite{ios-security}.  
    % No need to go this deep. The technologies used to protect PCC node are not relevant to the attack.
    % The OS has been trimmed down to a headless state that will only allow pre-signed PCC components to run. This is achieved with both secure boot and a feature called Restricted Execution Mode where what can be executable is allow-listed. PCC nodes work similar to Trusted Execution Environments and are verified by the end user via remote attestation. This allows a user to make sure that the PCC node it will be accessing is indeed trustworthy. The software running on the PCC is publicly available online for audit and the metadata of each release is logged onto a blockchain-like tamper-proof log that is also publicly available.
\end{packeditemize}

In short, PCC detaches the user verification and authentication, hides user identifiable information behind an OHTTP relay, and uses tightly controlled hardware and software to ensure the security and privacy guarantees.

\subsection{Privacy Pass}

In \ai, the authentication is based on Privacy Pass. Privacy Pass~\cite{privacy-pass} is an authentication scheme designed to detach the authentication process, which involves sensitive identifiable information, from using a service. Conventional authentication schemes offer an access token after verifying a user who will use the token to access the services. This means that a token is linked to a specific user which can be used for usage tracing, leading to privacy concerns. To solve this privacy issue, Privacy Pass was introduced~\cite{privacy-pass-paper}.

In Privacy Pass, a user sends in authentication information for eligibility verification along with an encrypted unsigned token. The authentication service, after verifying the information, signs the encrypted unsigned token using a blind signature based on Oblivious Pseudorandom Functions (OPRF)~\cite{privacy-pass-verification}. A blind signature allows the signer to sign a piece of encrypted data without the need to decrypt it~\cite{OPRF}. The user can then decrypt the now blindly-signed token and use it to access the service. For the service, while it can verify that this token is properly signed, since when performing the authentication it was in an encrypted form, it cannot know who the user actually is other than the token bearer is eligible to use the service.

\subsection{macOS Keychain}

Keychain~\cite{keychain} is a special database infrastructure on Apple *OS systems for the storage of passwords, access tokens and certificates. Keychain is available to all applications on the system with APIs for adding, deleting, retrieving and updating items given that both the application and the user are authorised to do so. There are four `keychain' databases used for different purposes: The `login keychain' is bound to a local user for local storage accessible only by that user. The `iCloud keychain' is bound to an Apple account and is synced across the specific user's devices. Any process accessing the iCloud keychain needs to designate a restricted entitlement that is signed by Apple~\cite{mac-kag,restricted-entitlements} and enforced by the AMFI in the kernel~\cite{amfi-exempt}. The `System keychain' is for items shared across the system which is readable by all users but modifiable only by sudoers. The `System Root keychain' is immutable and stores critical information such as root certificates. When an application tries to access a keychain item, an authentication prompt will pop up for the user to grant permission given that the user has access to that keychain item in the first place. Keychain is known to be a commonly targeted infrastructure with many known vulnerabilities~\cite{amos,CVE-2017-7150}.
\section{Authentication and Authorisation Workflow Anatomy}
\label{sec:workflow}

%Original
To offer a comprehensive understanding of the root cause of the \sysname attack, we present our detailed analysis of the authentication and authorisation process %\sx{I suggest to remove `authorisation'. Also, use British spelling. It's `authorisation' not `authorization'.} \hz{Just be safe. I think adding `authorisation' does no harm. Should find no more American spelling.} \sx{I don't find adding authorisation helps. It looks weird to me.} 
of \ai's request workflow. Our analysis is based on macOS 26.0 (25A353) `Tahoe' running on a Mac mini with an M4 chip.

\begin{figure}[t]
    \centering
    \includegraphics[width=0.85\linewidth]{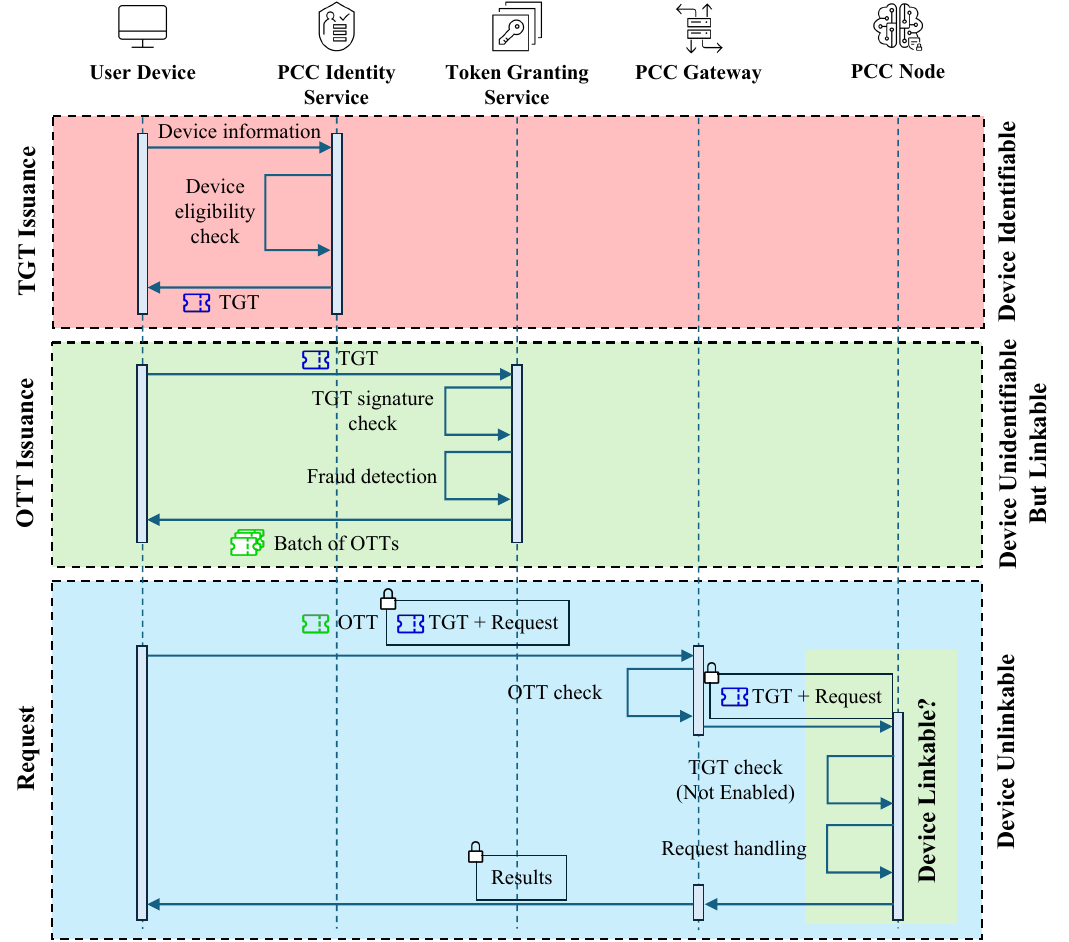}
    \caption{Authentication protocol of \ai}
    \label{fig:ai-protocol}
\end{figure}

\begin{figure}[t]
    \centering
    \includegraphics[width=.6\textwidth]{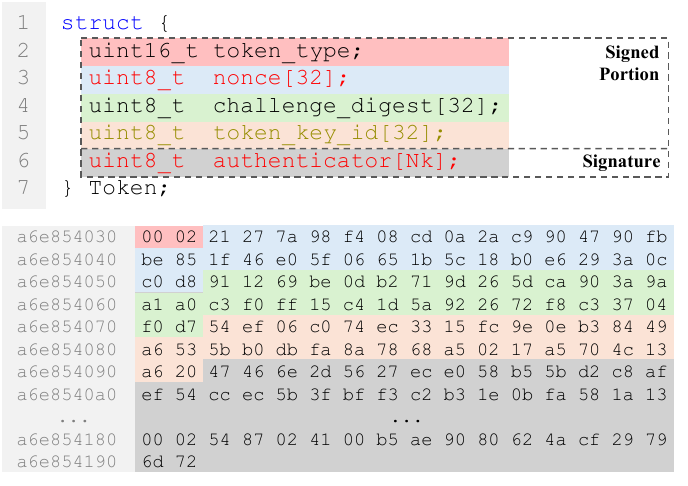}
    \caption{The token format of TGT and OTT following RFC 9578. The \textcolor{red}{red} fields are unique fields across different tokens. The \textcolor{olive}{green} field can be the same depending on the key selected. A colour-coded example of a TGT extracted from our Mac mini is provided below.} 
    \label{fig:token}
\end{figure}

\subsection{Device Eligibility and TGT Issuance}
\label{subsec:tgt}

Before a user can initiate a request to the \ai services in the cloud, it has to first be verified for its eligibility by proving to the PCC Identity Service that it is an authentic piece of Apple hardware running a proper OS. In this step, the user will obtain a Token Granting Token (TGT) from the PCC Identity Service which follows the RFC 9578 Privacy Pass protocol.\looseness=-1

The user that tries to access \ai will first create a request for TGT issuance. Per Apple's documentation~\cite{pccdoc}, Apple followed RFC 9578 using RFC 9474 RSA Blind Signature as the technical foundation for TGT. The format and a real TGT payload obtained from our traffic analysis are illustrated in \autoref{fig:token}. The TGT request contains token information that is packaged in a blinded message format and the device information. The token information is essentially the RFC 9578's challenge digest, nonce, and token type. By reverse engineering the \texttt{NetworkServiceProxy} private framework, we found that the challenge was a byte sequence obtained from the private API \texttt{NSPPrivateAccessTokenChallenge} which follows the Token Challenge Structure in RFC 9577~\cite{privacy-pass-scheme}, with the \texttt{issuer\_name} set to the TGT's issuer endpoint \texttt{tis.gateway.icloud.com} and the rest filled with 0. This challenge appears to be static \emph{so the only difference between each TGT (even across different devices) comes from the nonce portion of the TGT}.

% You have already said this in the first paragraph
%This request is sent to the PCC Identity Service which is a separate service outside of the PCC inference request handling workflow. According to Apple's documentation~\cite{pccdoc}, the PCC Identity Service will verify the device information and, if passed, sign the blinded message using a long-lived key. 

% Below interpretation and observation are too detailed.

%Through our reverse engineering, we found out that Apple also refer to this token as a Long-Lived Token (LLT) in their code. When the user retrieves the signed blinded message, it can unblind it to retrieve a properly signed TGT. The TGT is then stored on the device.

% This design means that the PCC Identity Service, while knowing that who requested a TGT, cannot know the exact content and identifiable information from the unsigned TGT (the nonce), and therefore cannot link the TGT to the actual device although the TGT is per-device. This also means that \ul{a TGT's redemption does not have a verification to whom it was issued to}.

% We observed that via empirical experiments is that while Apple's documentation claims that TGTs are issued on ``a per-user, per-device basis''~\cite{pccdoc}, different user accounts on the same Mac received the same TGT in our tests. Specifically, two distinct macOS user accounts logged into different Apple accounts on the same Mac mini both obtained identical TGTs. This also hints that the nonce portion of the TGT might be derived from device information.

\subsection{Service Authorisation And OTT Issuance}
\label{subsec:ott}

Once the user obtains the TGT, the rest of its interactions with the \ai platform no longer requires device eligibility validation. To achieve usage limitation, each request a user makes needs a One-Time Token (OTT) which is redeemed using the TGT from the Token Granting Service. Here, the user is hidden from the Token Granting Service via OHTTP relay. We suspect the reason for using another stage of token is that while TGT is not linked to the actual device, the TGT itself can be somehow treated as an identifier itself, causing privacy concern. By redeeming TGT into OTTs, these OTTs are now completely unlinked from a single client because they differ every time.

%Interestingly, OTTs' issuance also follows the same RFC 9578 protocol as in \autoref{fig:token} with the only difference this time to change the \texttt{issuer\_name} in the challenge to OTT's issuer \texttt{rts.gateway.icloud.com}. Besides, its nonce is derived from the TGT which we will discuss more details in \autoref{subsec:token-val}. The request is now sent with the TGT obtained from the last step. 
In practice, we found that OTTs are fetched in batches with a target amount of 50 OTTs cached in a user's device. The Token Granting Service signs the OTTs with an OTT key which has a shorter lifespan compared with TGT. OTTs are also stored on the device and the user can spend one token per request in the future. We can also observe that \ul{OTTs are also not bound to who they were issued to}.

Other than service authorisation, the OTT issuance process also serves as the rate limiter which manages each user's quota allowance. Our tests show that \ai has a quota limit of roughly several millions of tokens a day per user. Once the quota is used up, a user has to wait until roughly the next day before it can receive any new OTTs, essentially blocking the user from continuing to use the service. %From Apple's documentation, this is achieved by an 8-bit \textit{fraud report data} associated with a specific TGT. This fraud report data is used by another service called as Fraud Detection Service with very few details disclosed~\cite{pccdoc}. Since the 8-bit data is very small and is not device specific, we believe it can be regarded as part of the TGT.

\subsection{Authentication in User Request Handling}
\label{subsec:req-handle}

%After a user receives both TGT and OTTs, a user can issue GenAI requests to the \ai platform under the protection of the PCC scheme. There are also two stages for authentication in the user request handling, with one stage to be optional and currently disabled as suggested by Apple and the source code provided~\cite{pcc-source}.

% The user's request is packaged in two layers like an onion. The actual prompt payload is packed in an encrypted data body that only a PCC node can decrypt. This content is the most sensitive user-privacy-related data. Surprisingly, Apple also packed the TGT inside this payload. Outside of this encrypted body, Apple packaged the OTT and routing metadata together and formed the data structure to be sent to the \ai services.

\bheading{First Stage: OTT Validation on PCC Gateway.} After a user receives both TGT and OTTs, a user can issue GenAI requests to \ai under the protection of the PCC scheme. Even though the inside payload is already encrypted, Apple still uses TLS to encrypt the entire package including the OTT. This package is relayed via the OHTTP relay and will first arrive at the PCC Gateway, which is a gateway server that is not allowed to touch the user's data. PCC Gateway will decrypt the outer TLS layer and validate the OTT. The OTT is validated against the OTT's signing key. If passed, the OTT is considered as consumed and can't be used again, and the entire package (likely including the OTT, will be discussed below) is sent to the PCC node that handles the actual request.

\begin{figure}[t]
    \centering
    \includegraphics[width=.65\textwidth]{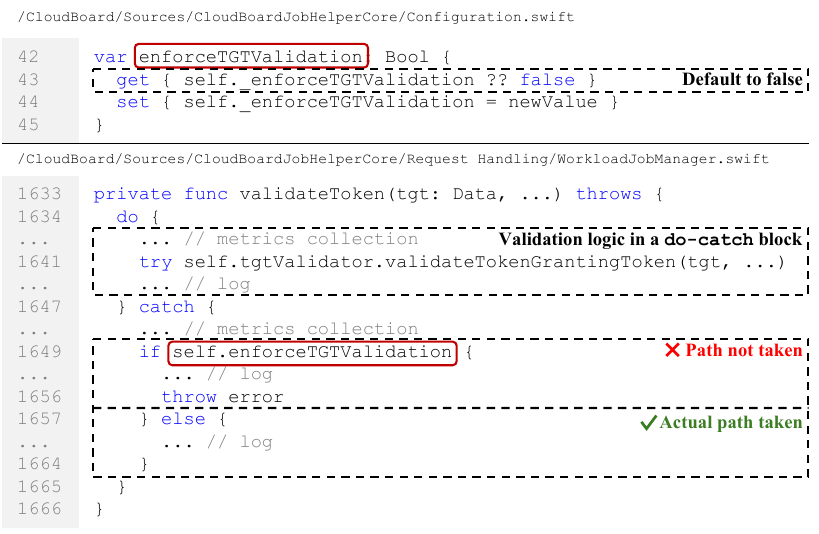}
    \caption{Swift code logics of CloudBoard in PCC showing the default result bypass behaviour of TGT validation.}
    \label{fig:tgt-bypass}
\end{figure}

\bheading{Second Stage: (Optional) TGT Validation on PCC Nodes.} On the PCC node that handles the request (or the PCC proxy node when multiple nodes are involved), the encrypted payload is decrypted to reveal the TGT and the user prompt. Here, an optional validation of the TGT can happen to verify that the TGT is valid. According to Apple's documentation~\cite{pccdoc}, this validation is not currently implemented but may be used in the future for abuse mitigations. However, from the source code provided, we actually found out that the verification has been implemented and is verified for each request as shown in \autoref{fig:tgt-bypass}. However, there is a configurable option called \texttt{enforceTGTValidation} that defaults to \texttt{false} that suppresses the error throwing, meaning that even if an invalid TGT is attached and the TGT validation fails, the error will not trigger an abortion in the processing.

\subsection{Token Validation and Expiry}
\label{subsec:token-val}

One of the critical procedures omitted from Apple's documentation is how the tokens are validated and how the expiry is determined. We summarise our discovery via reverse engineering, empirical tests and source code analysis. While the validation on Token Granting Service is not publicly documented, since the token validation is mostly cryptographic operations, we believe it follows a similar routine as the validation on PCC nodes.

\bheading{Token Validation.} The token validation process is similar to the RFC 9578's specification. In the method \texttt{validate} of \texttt{TokenGrantingTokenValidator} of the \texttt{CloudBoard} on PCC nodes, it first filters the keys in its key set that are currently still valid by the key's \texttt{validStart} and \texttt{validEnd} properties. From the filtered key set, the method will match the \texttt{tokenKeyID} field (known as the \texttt{token\_key\_id} in RFC 9578) with the valid keys. If a key matches, the key will be used to verify the signed portion against the signature (\texttt{authenticator}) of the token. If the signature is accepted, the token is considered as valid. For TGTs, since the nonce is randomly generated, a valid signature is considered as sufficient by Apple. While in \autoref{subsec:tgt} we observed that the nonce might be derived from device information, the nonce itself was not verified for any meaning during the validation process. For OTTs, an additional validation is needed.

%Left detailed analysis on OTT validation. TGT alone is enough.
%From the source code, we did not observe any checks to the \texttt{challenge\_digest} field. This is likely due to the fact that the challenge remains static and exists more like a feature to comply to RFC 9578.

%\bheading{Additional Validation for OTTs.} The nonce field in OTTs is not a simple random number. Instead, it is a value derived from the TGT. From the method \texttt{validateOttIsDerivedFromTGT}, we can observe that the nonce of an OTT is computed as a SHA256 hash of the TGT appended with a salt value. The salt value appears to be a random number generated by the Token Granting Service during OTT fetching. From Apple engineer's comments in the code, the intention of doing this is to ensure that \textit{the TGT was valid at the time of OTT issuance}. From the code analysis, it looks more like making sure that the TGT and OTT arriving in a PCC node to be paired.

\bheading{Token Expiry.} From the above analysis and the data structure in \autoref{fig:token}, we can easily observe that there is no token expiry information in the token itself. The expiry is, instead, enforced by the key used to sign the token. This means that unless keys are generated at all times and updates the expiry very frequently, the tokens are likely to expire in batch for the same key. This also means that there is no way for a client to revoke a valid token or even check when will a token expire. To test out how long the keys remain valid, we performed an empirical test. From our observation, TGTs appeared to survive for at least a scale of several days which might be the reason for their internal name as Long-Lived Tokens. OTTs were much shorter-lived and were generally expiring in about 12 hours. \looseness=-1

\subsection{On-Device Token Handling}

\begin{figure*}[th]
    \centering
    \includegraphics[width=\textwidth]{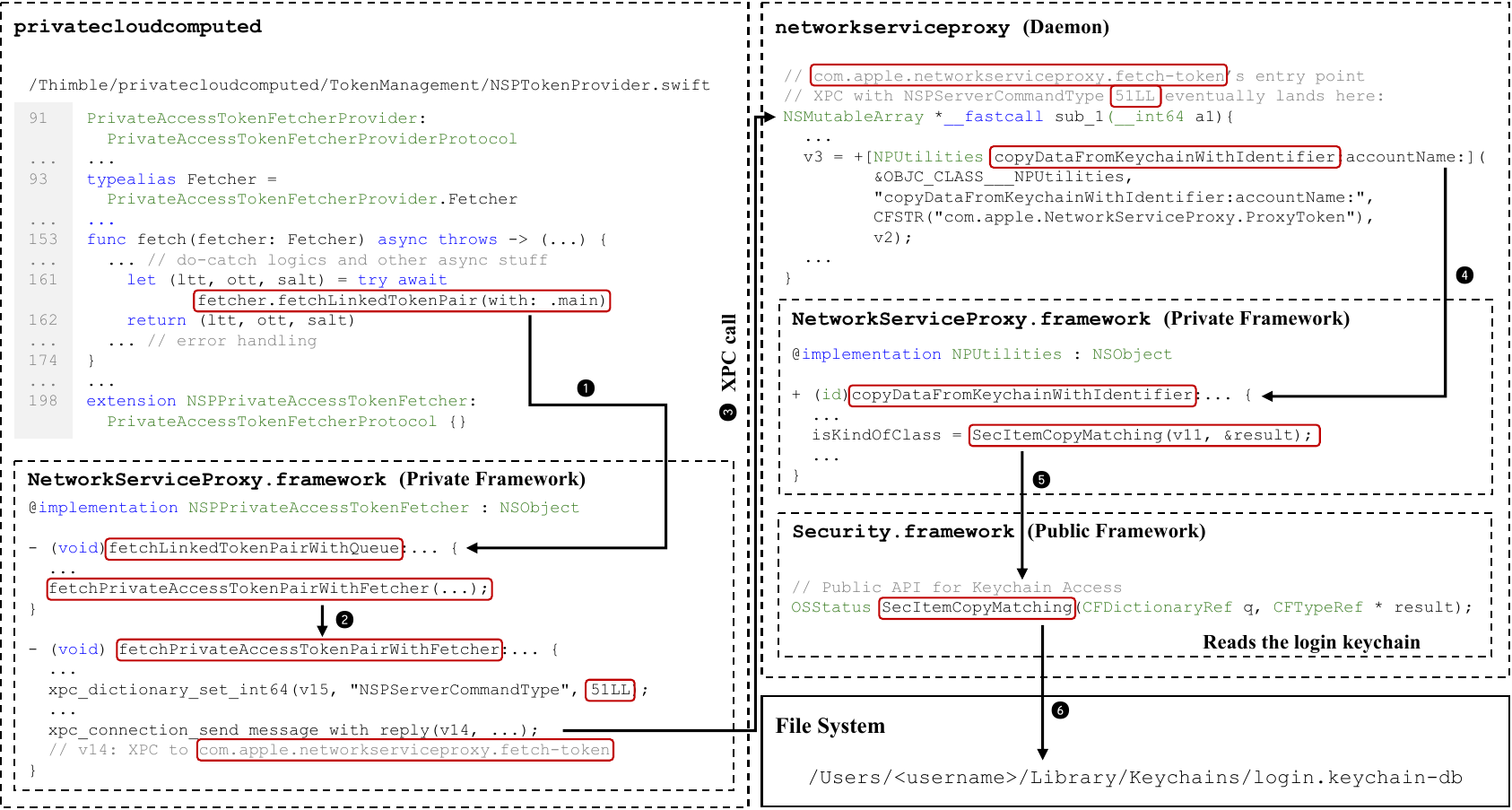}
    \caption{Workflow of the token fetching of \texttt{privatecloudcomputed}}
    \label{fig:keychain-re}
\end{figure*}

A key question here is where are the tokens stored on the device. Apple's public documentation does not offer any information on it. The source code published on GitHub, while including some clues, also does not point to the on-device storage of the token. To find out how the tokens are handled, we reverse engineered the \ai subsystem on macOS and found that multiple processes and frameworks are used. The workflow of retrieving the tokens is illustrated in \autoref{fig:keychain-re} and adding/updating the tokens works in a similar way. 

The published source code of the main daemon of \ai, \path{privatecloudcomputed}, uses an aliased class \texttt{Fetcher} which is an extension of the \path{NSPPrivateAccessTokenFetcher} that implementes the \path{fetchLinkedTokenPair} function to fetch the token (\ding{182}). \texttt{NSPPrivateAccessTokenFetcher} is implemented in Apple's private framework \path{NetworkServiceProxy}. From this point, no further public information is available and we performed reverse engineering. We found that the \texttt{fetchLinkedTokenPair} will call \path{fetchPrivateAccessTokenPairWithFetcher} for token fetching in an asynchronous way (\ding{183}) using an XPC~\cite{mac-xpc} call to initiate an inter-process request to the system daemon \texttt{networkserviceproxy} with the endpoint of \path{com.apple.networkserviceproxy.fetch-token} (\ding{184}). According to Apple's description on macOS's \texttt{man} manual, \texttt{networkserviceproxy} offers `\textit{Transparent network proxy for Apple system services}' and `\textit{The control plane of the NetworkServiceProxy framework runs inside of networkserviceproxy}'. While not documented publicly, the private framework and daemon of \texttt{networkserviceproxy} appears to be related to the iCloud Private Relay feature~\cite{nsp}. Given the extensive similarity of the OHTTP relay and certain authentication mechanisms, we believe that Apple added certain \ai authentication infrastructures into this private framework. In \texttt{networkserviceproxy}, the \path{com.apple.networkserviceproxy.fetch-token} endpoint is dispatched using an XPC parameter \texttt{NSPServerCommandType} which, for token fetching, is \texttt{51}. The \texttt{NetworkServiceProxy} private framework is again used for copying the token (\ding{185}) which would call the publicly available APIs such as  \texttt{SecItemCopyMatching} in the \texttt{Security} framework (\ding{186}). These APIs are used to store/fetch credentials in the keychain infrastructure. Our investigation showed that the tokens for \ai are stored in the login keychain (\ding{187}). This will serve as the extraction point of the tokens and the same APIs which are used in our attack will be introduced in later sections.\looseness=-1
\section{The \sysname Attack}
\label{sec:attack}

With a comprehensive understanding of the authentication workflow, we discuss the vulnerability in this authentication flow and present our attack, \sysname.

\subsection{What's Wrong with the Protocol?}
\label{subsec:4.1}

\begin{table}[t]
\caption{\ai access tokens and their fields}
\label{tab:tokens}
\centering
\small
\resizebox{\textwidth}{!}{
\begin{tabular}{l|llll|l|l}
\hline
\hline
\multirow{2}{*}{\textbf{Token Name}} & \multicolumn{4}{l|}{\textbf{Signed Portion}}                                                                                                                & \textbf{Signature} & \multirow{2}{*}{\textbf{Validity}}     \\ \cline{2-6} 
                                     & \multicolumn{1}{l|}{\texttt{token\_type}} & \multicolumn{1}{l|}{\texttt{nonce}} & \multicolumn{1}{l|}{\texttt{challenge\_digest}} & \texttt{token\_key\_id} & \texttt{authenticator} & \\ \hline
TGT                                  & \multicolumn{1}{l|}{Static}               & \multicolumn{1}{l|}{Unknown \& Unverified}         & \multicolumn{1}{l|}{Static}                     & Key-Related             & Key-Related & Days           \\ \hline
OTT                                  & \multicolumn{1}{l|}{Static}               & \multicolumn{1}{l|}{TGT + Random}     & \multicolumn{1}{l|}{Static}                     & Key-Related             & Key-Related & $\sim$12 Hours           \\ \hline
\end{tabular}}
\end{table}

%While the protocol may appear to be rock-solid while balancing privacy at a glance, we discovered that the design of the tokens and the handling of them suffers from several problems. To help better illustrate these problems, we summarised the fields of the tokens in \autoref{tab:tokens}.

\bheading{Lack of Client-Side Token Protection.} A problem with \ai's access tokens is that the tokens are pretty much open to anyone on the device without any special privilege requirement. In particular, the TGTs and OTTs are stored in the login keychain which is accessible with simply the current user's privilege. Besides, they are stored in plaintext in the keychain, meaning that an attacker has a common entry to extract the token without the need to resort to debugging the \ai processes like \texttt{privatecloudcomputed} daemon. In summary, an attacker has a simple and unified way to access the tokens with no privilege escalation required.

\bheading{No Strong Bound to Hardware.} Apple's privacy design, instead of just making the tokens not traceable to a specific device, completely detaches the tokens from the device. While we discovered that the nonce of a TGT might be related to device information, neither is it verified nor device information is sent to a validator (Token Granting Service, PCC gateways, PCC nodes). This means that in no way can a token validator verify that a request comes from the hardware that the token is issued to.\looseness=-1

\bheading{Long-Lived Token Can Be Leaked.} The storage scheme and protection (if it exists at all) is the same for TGTs and OTTs. Given their nature of single usage, leakage of OTTs might be less severe and can be detected with PCC node's currently disabled TGT validation. However, the leakage of TGTs is significantly concerning. The danger of TGT leakage comes from two aspects. First, a TGT's expiration cycle is at a scale of days. Second, a TGT can be used to redeem OTTs as long as the TGT is valid and the rate limit is not reached. Even if the rate limit was reached, after a cool down period, we observed that a TGT can again be used to redeem OTTs. This gives an attacker a very large window to access \ai services using the victim's actual allowance. In our experiments, the daily rate limit resets after a cooldown period of roughly one day, while a TGT remains valid for several days. This means that an attacker holding a stolen TGT can consume the victim's full daily quota, wait for the cooldown to pass, and then consume the quota again on subsequent days---effectively exhausting the victim's rate limit multiple times within a single TGT's lifetime.

\bheading{No Revocation Available.} As a remedy of last resort, token revocation may help stop a stolen token to be continuously used by an unauthorised attacker to prevent further loss. However, we can see that in \ai platform, a user has no way to revoke or even refresh its access tokens. This also means that if an attacker was able to obtain the TGT, it can keep on using the TGT with no concern that the token may be invalidated by the user.

In summary, \ul{Apple Intelligence's access tokens are poorly protected, insufficiently validated, and can be stolen for long-term use by an attacker with no remedy available to the user}.

\subsection{\sysname: Attack Against \ai's access tokens}

With the vulnerabilities in mind, we now introduce our attack, \sysname, which can allow %\sx{Debate `allows' and `can allow'} \hz{? What is worth debating here? `can allow' is not English. Either `Serpent allows' or `Serpent enables'} \sx{Why can allow is not English? I used `can allow' to tone down. Remember that \sysname requires user interaction, meaning that \sysname alone is not enough. When I say `can allow' it means similarly to `may allow' or `might allow'} 
a malicious attacker to extract the tokens from a victim's Apple device and use them on their own devices. The attack contains two phases: An \textit{extraction phase} on the victim's device, and a \textit{disguise phase} on an attacker-controlled machine.

\bheading{Threat Model.} We consider the victim to be an ordinary owner of a Mac. Like most users, the victim runs normal applications downloaded from the Internet at regular user privilege. In \sysname, we assume the victim downloads a malware distributed by the attacker and runs it on its device. When the malware attempts to read the access tokens from the keychain, macOS presents an authentication prompt requiring the user to click `Allow'. We assume the victim grants this permission, an assumption we consider realistic given the frequency of similar prompts in normal macOS usage, the non-technical nature of the prompt text, and community advice that commonly recommends clicking `Allow' on such prompts~\cite{keychain-pwd-1,keychain-pwd-2,keychain-pwd-3}. In \autoref{subsec:victim-aware}, we present a detailed analysis to support this assumption. %\sx{Refer to later section 5.2 here}. 
The attacker sets up a remote server waiting for the malicious logic to send back the access tokens. If the access tokens were sent back and were usable for \ai services, the attack is considered as successful.

\bheading{Extraction Phase.} In the extraction phase, the victim runs an app obtained from the Internet that contains the attacker's malicious logic. This logic is very simple and can be done either programmatically or via a simple shell script. When done programmatically, the app can simply call keychain API \texttt{SecItemCopyMatching} with the token's information. 
% Left the specific names
%The \textit{account names} for the tokens are \path{tis.gateway.icloud.com} for TGTs and \path{rts.gateway.icloud.com} for OTTs. The \textit{service names} for the tokens are \path{com.apple.NetworkServiceProxy.PrivateAccessTokens.LongLivedTokens} for TGTs and \path{com.apple.NetworkServiceProxy.PrivateAccessTokens.OneTimeTokens} for OTTs. 
If the attacker chooses to use a shell script, the \texttt{/usr/bin/security} can also be used with the same account names and service names to fetch the tokens. The result from this step is a property list format (also known as a \texttt{plist}). The actual token is encoded as Base64 in the \texttt{plist} at the path of \path{$objects/NS.data}. The rest of the \texttt{plist} contains simple metadata that is not useful to an attacker. When retrieving the tokens from the keychain, it may require a user authentication if the app was never authorised to access the keychain. This authentication does not need root privilege and is common for any app that uses the keychain. Once the tokens are extracted, the malicious logic can now send them to the attacker in a remote place.

\bheading{Disguise Phase.} In the disguise phase, the attacker will use the tokens obtained from the victim to disguise itself as the victim. This step is surprisingly simple. On the attacker's Mac, the attacker can overwrite the tokens of its own with the victim's tokens. If no such tokens exist yet, the attacker can create these keychain entries with the same account and service names and with the tokens in the same \texttt{plist} format as the keychain item's password payload. Now, whenever the attacker's Mac issues a request to the \ai services, it will use the tokens of the victim. When the fetched OTTs are exhausted, the TGT can also be used to fetch a new batch of OTTs as if the fetch request was initiated by the victim. \looseness=-1

\section{Security Analysis}
\label{sec:eval}

In this section, we systematically discuss and analyse the \sysname attack. In particular, we discuss and evaluate the effectiveness of the attack, the victim experience during and after the attack, the implications of this attack, and the possibilities for other abuses enabled by the attack. We performed our empirical experiments on the same Mac mini on macOS 26.0 (25A353). 

\subsection{Effectiveness of \sysname}
\label{subsec:effectiveness}

We evaluated the effectiveness of \sysname in three ways: First, we perform traffic analysis for theoretical feasibility; Next, we use valid tokens to revive a banned Mac for a practical unauthorised access; And finally, we present a DoS attack to demonstrate that the usage was counted towards the victim.

\bheading{Nominal Owner of Request.} The very first thing we need to confirm is that after the successful import of the victim's tokens, the attacker is \ul{indeed using the victim's token} instead of still using its own. To test it, we intercepted the requests sent by the \texttt{privatecloudcomputed} after importing the victim's tokens. We observed from the traffic that the TGT and OTTs included in the requests were indeed from the victim. This proved that the nominal owner of the request is the victim.

\bheading{Reviving Ineligible Macs.} To demonstrate a concrete attack and to further verify that no other ownership information was used in the \ai's authentication flow, we performed the experiment of reviving an ineligible Mac owned by the attacker. 

To do this, we first sent a huge prompt with a size of millions of prompt tokens to exhaust the allowance of the attacker's Mac. This will essentially cause the Mac to be banned from \ai services temporarily for at least a day. We then deleted the TGT and repeatedly fetched new TGTs until PCC Identity Service also banned the Mac. This circumstance represents an abuse scenario of \ai service by sending out a massive amount of requests. Now, the Mac was banned from the entire two stages of \ai's authentication flow and was ineligible to use \ai services. Next, we imported the access tokens obtained from the victim's device. After this, we found out that the attacker's Mac was now able to use the TGT and OTTs without a problem to access \ai services. The TGT was also fully usable to fetch new OTTs. This proved that the authentication to \ai services is entirely based on the access tokens of TGT and OTTs and our attack holds a very strong effectiveness in disguising the attacker to use the victim's \ai allowance.

\bheading{DoS Attack Against a Victim.} To prove that usages of the attacker using the stolen tokens are indeed counted towards the victim's, we also present a Denial-of-Service (DoS) attack against a victim user. In this case, the attacker will use up all the allowance of the victim by constantly redeeming for OTTs, essentially burning the tokens out. Our experiments showed that after the attacker used up the allowance of the TGT on the attacker's Mac, the victim's Mac also displayed a warning suggesting that `\ai is currently not available' and stopped, meaning that the usages were indeed counted towards the victim's allowance. As in our observations discussed in \autoref{subsec:ott}, Apple has rate limits for \ai service usage using their `Fraud Detection Service' mechanism by controlling how many OTTs can be redeemed from a TGT. This means that the attacker does not even have to actually use the OTT for the DoS attack. All it has to do is to constantly redeem for OTTs using the TGT and drop them.

% \begin{wrapfigure}{r}{0.5\textwidth}
%   \begin{center}
%     \includegraphics[width=0.48\textwidth]{birds}
%   \end{center}
%   \caption{Birds}
% \end{wrapfigure}
% \begin{figure}[t]
%     \centering
%     \includegraphics[width=.6\textwidth]{figures/keychain_pwd.png}
%     \caption{Keychain authentication prompt that pops up when an attacker tries to steal the tokens on macOS 26.0.}
%     \label{fig:keychain-pwd}
% \end{figure}

\subsection{Victim Awareness}
\label{subsec:victim-aware}

A key question to the \sysname attack is how it would visibly impact the victim during and after the attack. 

\bheading{During the Attack.} Since the attack extracts the tokens from the keychain, the victim has to authorise the access. We consider three ways to achieve keychain access:\looseness=-1
\begin{packeditemize}
    \item \textbf{Just Ask in Plain.} The victim will be prompted by the keychain subsystem of macOS to authenticate for the keychain item. %\st{The authentication window looks like the one shown in %\autoref{fig:keychain-pwd}} \hz{I deleted the prompt window figure to 1) save space and 2) draw people's attention away from the strong attacker model}. 
    The prompt shows the app's name and the service name of the TGT or OTT. Note that app's name can be customised by the developer. We believe that for an ordinary consumer, the prompt on the window would be difficult to understand and a consumer would seek help from the Internet. From discussions of Apple's own community, for a similar prompt, the general suggestion to handle this window is to type in the login password and hit the Allow button~\cite{keychain-pwd-1,keychain-pwd-2}. From other Internet forums, there are even suggestions to hit the Always Allow button~\cite{keychain-pwd-3}. In conclusion, we believe that there is a high chance of a victim to allow the malicious logic to access the tokens.
    \item \textbf{Fake an Authentication Prompt.} If the attacker would not want to display the default keychain prompt that may expose the attacker's intention, it can use a fake authentication prompt that appears to be innocent to obtain the victim's keychain password and then use tools with headless modes like \path{/usr/bin/security} to gain access to the tokens. This kind of attack has been observed in the wild in malware such as The Atomic macOS Stealer (AMOS)~\cite{amos}.
    \item \textbf{Combine with a keychain Bypass Attack.} %When combined with a keychain authentication bypass attack, 
    In this case, the attack is essentially transparent to the victim with no victim awareness at all. For example, attacks such as CVE-2017-7150~\cite{CVE-2017-7150} can bypass the authentication window, granting the malicious logics access to the tokens completely in the background without victim knowledge.
\end{packeditemize}

Once the attacker is authorised to access the tokens from the keychain, there will be no obvious or visible impact to the victim any more since all the attacker needs to do is to send the tokens back to the attacker via network.

\bheading{After the Attack.} After the attacker gets the access tokens, it can use it completely without the knowledge of the victim. If the attacker uses the tokens with some control by not completely depleting the allowance, the victim might not even notice any difference as long as the total requests of it combined with those issued by the attacker do not exceed the allowance. However, since the attacker is not traceable on the victim's side, if the attacker uses up the entire allowance the victim has, all the victim will find out is that it cannot use \ai services any more. As presented in \autoref{subsec:effectiveness}, there is no indication that the allowance has been exhausted but instead a warning saying `\ai is currently not available'. For an ordinary consumer, this looks more like a service interruption rather than that it has been attacked. Therefore, after the attack, the victim might experience usage difficulties or DoS symptoms but would not be aware that it has been attacked.

\subsection{Implications of \sysname}

The implications of the \sysname attack can be wide-spread given how many Apple devices are used in real life. Here, we introduce three main implications of \sysname:
\begin{packeditemize}
    \item \textbf{No Privilege Escalation Required.} In this case, the attacker must first achieve code execution on the victim's machine (e.g., through a malware). The entire token extraction operates at normal user privilege. The \sysname attack requires only normal user privilege, meaning that an attacker does not need to seek a complex attack chain with privilege escalation. As we have discussed in \autoref{subsec:victim-aware}, the keychain authentication prompt can be addressed through social engineering with a high chance that the victim would authenticate.
    % This repeats the intro's point about non-technical victims.
     %\item \textbf{Targets at Ordinary Consumers.} The \sysname attack targets at ordinary Mac users individually who rely on Apple's platform security design to defeat attacks instead of being a professional on cybersecurity on their own. Unlike API keys in the apps that are owned and protected by app developers, one cannot anticipate an ordinary user to be knowledgable on protecting these tokens.
     \item \textbf{Easy to Scale.} The methodology of \sysname is extremely easy to scale due to the unified attack surface of Apple devices. An attacker can easily insert the simple malicious logic in the apps they distributed to collect these access tokens from Mac devices around the world to form a token pool. The use of the tokens are also very easy to scale. An attacker can simply build a farm of the most affordable Mac minis to run requests 24/7. For any Mac that exhausted its token's allowance, that Mac can just pick up a new token from the token pool and immediately get back online.
     \item \textbf{Hard to Trace.} The end user's information (e.g., IP addresses, hardware identifiers) are hidden from \ai services by design. This means that even if an abuse has been identified, there is no possible way to know any information about the attacker on the Apple side. On the client side, while the token pool server's entry point may be exposed, such entry point can also be hidden through proxies to completely hide the attacker from anyone.
\end{packeditemize}

\section{Countermeasures}
\label{sec:defence} 

In this section, we present countermeasures we designed against the \sysname attack, one of which has been deployed by Apple. Due to the complex and non-standard nature of \ai's authentication and authorisation flow, a complete defence would inevitably need heavy changes to the protocol. We first offer two mitigations that, while they cannot completely fix the vulnerability, can make \sysname significantly more difficult while requiring minimal changes to existing protocols. We then offer three complete defence methods that fix the fundamental flaw of \ai's authentication flow at a cost of heavy modification to the flow and the software stack.

\subsection{De-privilege Tokens from Users} 
\label{subsec:depriv}

A critical reason that the \sysname attack works is because the access tokens are accessible to a regular user at normal user privileges. Therefore, the first possible mitigation is to prevent \sysname through the token storage access control. The current implementation stores them in the login keychain, allowing a user to easily create, read, update and delete all keychain entries which unfortunately include the tokens used by \ai services. An attacker in this case can easily query the entry for extraction and add/update the entry on its own machine for unauthorised access.

A relatively quick and easy mitigation to this problem is to simply de-privilege a regular user from accessing the tokens. To do that, a possible solution is to have a dedicated account for the \ai's client-side services. For example, a \texttt{pcc} account can be created to run the \texttt{privatecloudcomputed} daemon. The access tokens for users can be stored in this \texttt{pcc} account's keychain so that they are never available directly to a user. The benefit of this mitigation is that it requires almost no modification to the authentication protocol itself, with the only effort required to be a new system user. The downside is, however, that this is bypassable with root privilege and therefore is only a mitigation and not a fundamental fix. While being more difficult, a sudo prompt can be faked or misleading in a similar fashion as discussed in \autoref{subsec:victim-aware}.

\begin{figure}[t]
    \centering
    \includegraphics[width=.6\textwidth]{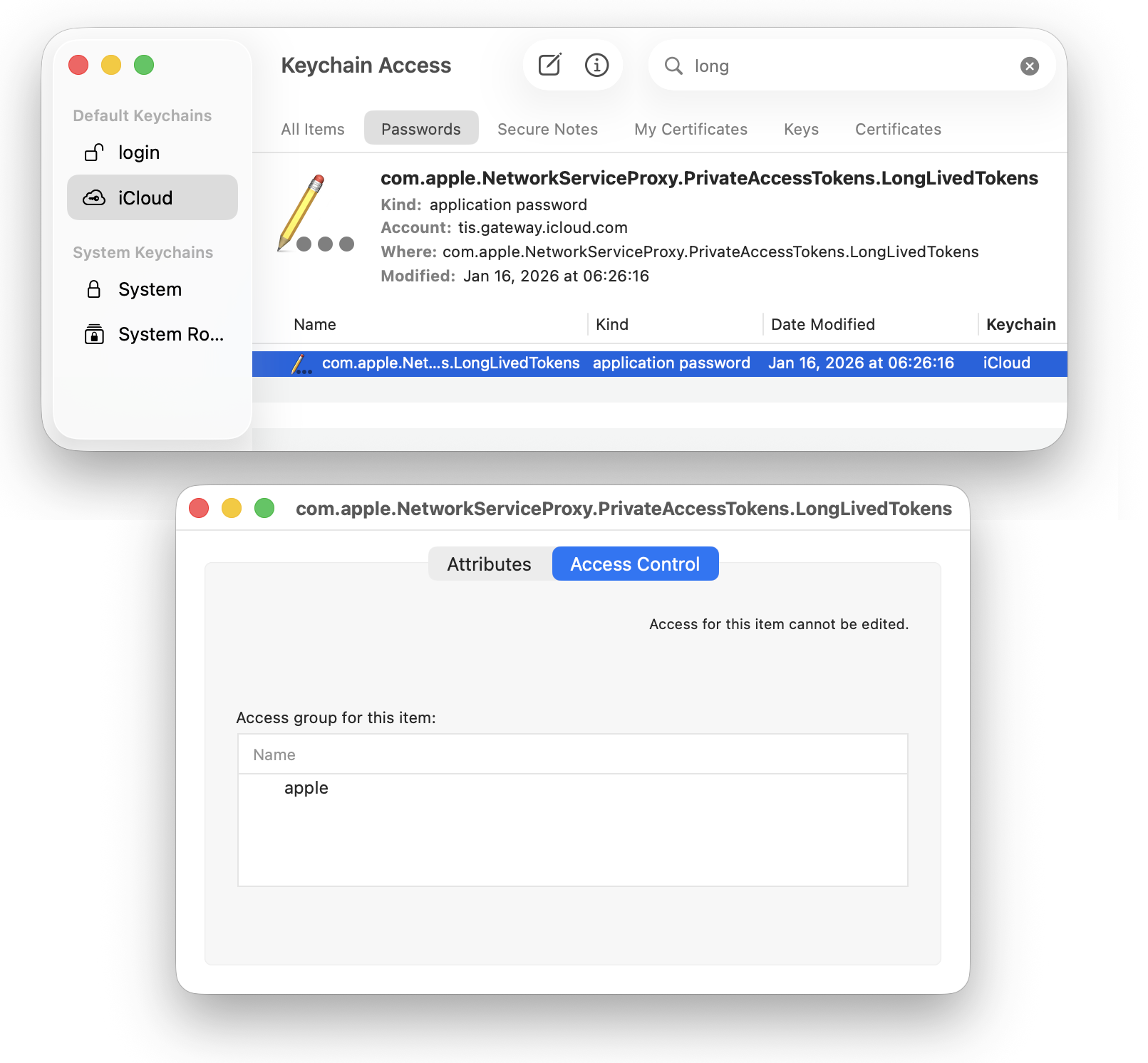}
    \caption{TGT now saves to the iCloud keychain on macOS 26.2 after Apple deployed the mitigation with access group set to `apple'.}
    \label{fig:failed-fix}
\end{figure}

\bheading{Mitigation Adopted by Apple.} We offered this insight to Apple along with others. After Apple claimed they have deployed a mitigation for \sysname in macOS 26.2 security update, we analysed their changes and found out that they likely have followed this idea but using another approach. Instead of separating a local user, Apple moved the tokens from the user's login keychain to the iCloud keychain with access group set to `apple' as shown in \autoref{fig:failed-fix}. Any applications or even CLI binaries trying to access the iCloud keychain must have a restricted entitlement designating matching \texttt{keychain-access-groups} identifiers~\cite{mac-kag}, which is enforced by AMFI~\cite{amfi-exempt} in the kernel. These restricted entitlements must be approved and signed by Apple for a special provisioning profile~\cite{restricted-entitlements}. This means that one would need to bypass this entitlement check if using the keychain APIs, making it more difficult. However, this mitigation is not a solid fix and may still be vulnerable to other methods. For example, \cite{amfi-exempt} offers a kernel extension to selectively bypass the restricted entitlement check. Furthermore, automated memory debugging and traffic analysis can still reveal the tokens used in \ai.
We have reported this to Apple and received its confirmation. Apple is currently working on new patches for this issue.

\subsection{Reducing TGT Signing Key Lifetime}
\label{subsec:reduce-tgt-lifetime}

%\sx{Where did this come from?} \hz{Added on Mar 30th, before your warning. Forgot this was AI generated. Will delete.} \sx{You can also rewrite the content by hand if it fits in. The concept should technically work.}
An intuitive server-side mitigation that Apple could employ without any client-side changes is to reduce the lifetime of the keys used to sign TGTs. As discussed in \autoref{subsec:token-val}, tokens expire when their signing key expires, and our experiments showed that TGTs remain valid for several days. By rotating TGT signing keys more frequently, for example, from a lifetime of days to hours, Apple could significantly shrink the window during which stolen TGTs remain usable. %--- \sx{These dashes look pretty LLM to me} 
Although this would not eliminate the attack, the attacker's ability to exhaust the victim's rate limit across multiple days would be limited. The trade-off is increased load on the PCC Identity Service, as devices would need to re-authenticate and obtain new TGTs more frequently. However, since TGT issuance is already a lightweight operation relative to inference, this overhead may be acceptable. This mitigation is complementary to the other countermeasures discussed below.

\subsection{No Plaintext Retrievable Storage} 

While keychain is an encrypted database, its APIs allow plaintext retrieval of the stored secrets including the access tokens for \ai. Our second proposed mitigation aims to make it significantly more difficult to steal the tokens by not storing retrievable plaintext of the tokens. The main idea is: The only moment the tokens are in their plaintext form should be in the memory of the PCC daemon. There are two possible ways of achieving this: \looseness=-1
\begin{packeditemize}
    \item \textbf{Encrypt Using Hardware Root-of-Trust.} The tokens can be encrypted using hardware root-of-trust such as the Secure Enclave Processor (SEP)~\cite{sep} that is trusted, verifiable and isolated from the rest of the OS. A new API can be implemented for SEP that would only decrypt a token requested from the PCC daemon.
    \item \textbf{Request New TGTs Every Time.} Given that PCC daemon does not typically exit, an even simpler solution is to request new TGTs whenever PCC daemon launches. This means that there will never be any local storage of the tokens at all. There can be slightly increased traffic for the PCC Identity Service.
\end{packeditemize}
In this way, since there is no local retrievable storage of the tokens in plaintext, an attacker must be able to attach a debugger onto the PCC daemon to hunt for the tokens. With widely-used and default-on security techniques like ASLR, the difficulty of automated harvesting tokens can be significantly increased. The benefit of this approach is that it also requires little to no modification to the protocol but only changes to the OS components. However, just like the previous approach, it is a mitigation and an attacker, albeit difficult, can still extract the tokens.

\section{Discussions}

%\subsection{Other Abuses Enabled by \sysname}

The \sysname attack assumes a threat model under which an attacker steals the \ai access tokens to use the services while counting the usage towards the victim. However, the vulnerability that enabled \sysname can be exploited for further abuses, and could be used directly against Apple.

One of the interesting observations we found about \ai is that none of the request handling processes after the initial PCC Identity Service need, or at least, verify an Apple device. \ai services are designed to be used with interfaces (mostly GUI) designed by Apple. These interfaces are generally limited with predefined features, and are designed to be used by a real person instead of being a generic LLM service. Due to the authentication flow vulnerability used by \sysname, we believe that it is possible to create a fully-automated compatible client that would work on other platforms, such as Linux, that only needs the TGT (and optionally, OTTs) and provide low-level customisations to the prompt of the requests  with the open-source client-side code on GitHub and other publicly available knowledge. Unlike the human-oriented interfaces of \ai, the client can provide an automated interface to use \ai services as a generic LLM service which could lead to a significantly higher amount of requests than that from an ordinary user since an automated client can easily use up all the rate limit in a short amount of time. Even worse, this makes reselling the \ai services as a generic LLM service possible and likely profitable, even at the cost of an entry-level device with \ai capabilities such as the Mac mini. In this case, an attacker may even purchase these devices out of its own pocket for this level of abuse.\looseness=-1

\paragraph{Ethics and Responsible Disclosure} We have responsibly disclosed the vulnerabilities to Apple which confirmed the attack, awarded the bug bounty, assigned CVE to it, and released a software update (macOS 26.2) with patches addressing the vulnerability. Our attack experiments were performed exclusively on the devices we own. We did not exploit public-facing services Apple offers nor did we interfere or compromise normal users.  

\section{Related Works}
\label{sec:related}

%\sx{Make a table to discuss}

\begin{table}[b]
\centering
\caption{Comparison of \sysname with other attacks}
\label{tab:related-comparison}
\resizebox{\textwidth}{!}{
\begin{tabular}{rr|l|l|l|l|l|l}
\hline\hline
\multicolumn{2}{r|}{\textbf{Attack}}                                              & \textbf{Victim} & \textbf{Victim Awareness} & \textbf{Valid Length} & \textbf{Detection} & \textbf{Technique} & \textbf{Remedy}     \\ \hline
\multicolumn{2}{r|}{\textbf{\sysname}}                             & End User            & Low                & Long (TGT) & None               & Extraction            & None       \\ \hline
\multicolumn{1}{r|}{\multirow{2}{*}{\textbf{API Key Leakage}}} & \textit{Active}  & Developer             & High               & Long & Usage              & Analysis           & Revocation \\ \cline{2-8} 
\multicolumn{1}{r|}{}                                          & \textit{Passive} & Developer             & High               & Long & Usage              & Scan               & Revocation \\ \hline
\multicolumn{2}{r|}{\textbf{Cookie Theft}}                                      & End User            & High               & Varies & Geo/Usage          & Varies            & Logout     \\ \hline
\end{tabular}}
\end{table}

In this section, we discuss and compare research that are closely related to \sysname. \sysname is a credential theft attack, in which existing works can be primarily categorised into two types: API key leakage and cookie theft attacks. We provide a summarised comparison in \autoref{tab:related-comparison} and discuss them in details below.

%\subsection{API Key Leakage}

% \sx{API keys are app-specific. Generally stored in the ciphertext until app is launched. When app launched, plaintext is only in memory so need a debugger. Different apps handles API keys dramatically differently unlike Apple has a unified interface across its devices (Keychain). API keys are owned by developers. Developers are considered as cybersecurity professionals. API keys can be used to access services as allows by the key's issuer. Compare with ours.}

\subsection{API Key Leakage}

API keys are app-specific credentials that grant an app to access certain services used by the app (e.g., map for a food delivery app). The leakage of an API key means an attacker can use the key for its own interest while making the usage billed towards the victim app. Existing works have identified many possible leakage paths of these API keys. 
Works like \cite{appsecret,mini-apps,android-leak,ios-leak} used active methods including traffic analysis and reverse engineering to obtain the API keys that are inadequately protected across different platforms. Other works like \cite{sinha,secrethunter,lounici,saha} used a passive approach that scans for leaked API keys in opensource projects on platforms like GitHub. \sysname bears fundamental differences from these works in terms of both our approach and the attack's consequences. 

First, \ai's tokens are device and user specific, meaning that the victim, instead of being a supposed-to-be professional developer~\cite{api-leak}, is likely a non-technical end user. This means that the victim of \sysname should not be considered to be responsible for or even capable of properly protecting the tokens. 

Second, awareness wise, \sysname is significantly more stealthy. With proper usage alerts setup, a developer can quickly be aware of a potential API key leakage to take actions correspondingly~\cite{api-leak-detect-1,api-leak-detect-2}. However, as discussed in \autoref{subsec:victim-aware}, the victim of \sysname is almost impossible to tell if it is under attack. Even after the quota is used up, the seemingly service outage reveals no sign of \sysname. 

Third, instead of a single global secret on an app that the attacker can analyse locally pretending as a regular user~\cite{api-leak}, \sysname is an attack that harvests tokens from devices owned by non-specific victim users and could cause wide-spread issues. %An interesting property of \sysname is: Despite the attack aims at harvesting tokens from multiple or even a massive amount of victim users, as a built-in service, \ai's access tokens are stored in the same place across devices, meaning that the attacker does not need to address problems specifically for each individual victim. \looseness=-1

Finally, API keys can generally be revoked by developers in case of leakage to prevent further loss~\cite{api-revoke} but not for \ai. The victim of \sysname has no effective remedy available once the tokens are leaked, even if the victim becomes aware of potential leakages of its tokens. The only thing a victim can do is to wait until the long-term TGT gets expired. \looseness=-1

\subsection{Cookie Theft Attack}

% Cookie injection attacks
Cookie theft attacks are also related to \sysname. Cookies are credentials issued by a web service to an authenticated user to access certain services. Similarly to API key leakage, cookie theft attacks allow an attacker to access services but this time, with usages counted towards the victim user instead of an app. Given the nature of the standard yet insecure usage of cookies, many attacks have been introduced with different mechanisms such as XSS and Cross Site Request Forgery~\cite{csrf,csrf-2}, session hijacking~\cite{session-hijacking}, security policy forgery~\cite{sso-theft,cookie-theft}, DNS hijacking~\cite{cookie-pharming} and many more. However, when comparing with cookie theft attacks, while both targeting end users, \sysname also differs significantly.

Just like API key leakage, cookie theft can be detected both by the platform and the user relatively easily since cookies can leak information such as geolocation of IP address~\cite{ip-login-detect} and usage pattern~\cite{usage-pattern}. When a potential leakage is detected, the victim can be notified for awareness. However, \ai's authentication scheme is anonymous by design and has multiple stages to prevent revealing geolocations and linking usage to a specific user, meaning that there is no practical way to detect such leakage for user awareness.

%Besides, unlike the long-lived access tokens that serve as the only authentication credentials in \ai, cookies are generally shorter-lived and are combined with other stronger authentication methods such as two-factor authentications~\cite{two-factor-1} in case of potential fraud. This means that even if cookies get leaked, if the attacker is performing unusual behaviours such as high frequency usage of the service, the attacker may soon hit these additional authentications. However, for \ai, it is impossible to defend \sysname through this.

%Finally, 
Remedy wise, similar to API keys, most services offer a `sign out everywhere' feature that essentially invalidates all cookies previously issued~\cite{sign-out-everywhere} when the user feels that its cookies may have been leaked. However, for \ai, the fact that no revocation is available means that the victim will have no remedy once the tokens are stolen.
\section{Conclusion}
\label{sec:conclusion}

We have presented \sysname, the first practical attack against the access tokens of \ai. Our study systematically investigated the authentication and authorisation flow of \ai and discovered its vulnerabilities that could allow an attacker to easily extract the access tokens from a victim and use the service while counting the usage towards the victim. \sysname highlighted that access tokens need to be strongly bound to the device to ensure that they are only usable by the rightful user. 

\cleardoublepage
\appendix

\section*{Acknowledgement}
This research was supported in part by NSF awards 2112471 and 2207202. Any opinions, findings, conclusions or recommendations expressed in this material are those of the authors and do not necessarily reflect the views of NSF.

% optional clearing of the page
\cleardoublepage
\bibliographystyle{plainurl}
\bibliography{reference.bib}

@misc{apple-intelligence,
    title = {{Apple Intelligence - Apple}},
    author = {Apple},
    howpublished = {\url{https://www.apple.com/apple-intelligence/}},
    year = {2024}
}

@misc{pcc,
    title = {{Private Cloud Compute: A new frontier for AI privacy in the cloud - Apple Security Research}},
    author = {Apple Security Research},
    howpublished = {\url{https://security.apple.com/blog/private-cloud-compute/}},
    year = {2024}
}

@misc{pccdoc,
    title = {{Private Cloud Compute Security Guide | Documentation}},
    author = {Apple Security Research},
    howpublished = {\url{https://security.apple.com/documentation/private-cloud-compute}},
    year = {2024}
}

@misc{openai-auth,
    title = {{API Reference - OpenAI API}},
    author = {OpenAI},
    howpublished = {\url{https://platform.openai.com/docs/api-reference/authentication}}
}

@misc{mac-xpc,
    title = {{XPC | Apple Developer Documentation}},
    author = {Apple},
    howpublished = {\url{https://developer.apple.com/documentation/xpc}}
}

@misc{mac-kag,
    title = {{Keychain Access Groups Entitlement | Apple Developer Documentation}},
    author = {Apple},
    howpublished = {\url{https://developer.apple.com/documentation/bundleresources/entitlements/keychain-access-groups}}
}

@misc{nsp,
    title = {{Transparent network proxy for Apple system services wants to use the "login" keychain - Ask Different}},
    author = {Ask Different},
    howpublished = {\url{https://apple.stackexchange.com/questions/457623/transparent-network-proxy-for-apple-system-services-wants-to-use-the-login-key}}
}

@inproceedings{appsecret,
    author = {Zhang, Yue and Yang, Yuqing and Lin, Zhiqiang},
    title = {{Don't Leak Your Keys: Understanding, Measuring, and Exploiting the AppSecret Leaks in Mini-Programs}},
    year = {2023},
    isbn = {9798400700507},
    publisher = {Association for Computing Machinery},
    url = {https://doi.org/10.1145/3576915.3616591},
    doi = {10.1145/3576915.3616591},
    booktitle = {Proceedings of the 2023 ACM SIGSAC Conference on Computer and Communications Security},
    pages = {2411–2425},
    numpages = {15},
    address = {Copenhagen, Denmark},
    series = {CCS '23}
}

@misc{github-openai-key,
    title = {{I just searched ``OPENAI\_API\_KEY'' on GitHub and found thousands of exposed credentials}},
    author = {Albert John Lastima},
    howpublished = {\url{https://www.linkedin.com/posts/albert-maquiling-784612138_cybersecurity-openai-apikeys-activity-7345806283880480769-G_DL/}}
}

@inproceedings{api-leak,
    author = {Zuo, Chaoshun and Lin, Zhiqiang and Zhang, Yinqian},
    booktitle = {2019 IEEE Symposium on Security and Privacy (SP)}, 
    title = {{Why Does Your Data Leak? Uncovering the Data Leakage in Cloud from Mobile Apps}},
    year = {2019},
    pages = {1296-1310},
    doi = {10.1109/SP.2019.00009},
    address = {San Francisco, CA},
}

@INPROCEEDINGS{api-leak-detect-1,
    author={MengShanshan and Yang Xiaohui and Song Yubo and ZhuKelong and Chen Fei},
    booktitle={International Conference on Cyberspace Technology (CCT 2014)}, 
    title={Android's sensitive data leakage detection based on API monitoring}, 
    year={2014},
    volume={},
    number={},
    pages={1-4},
    doi={10.1049/cp.2014.1340}
}

@INPROCEEDINGS{api-leak-detect-2,
    author={Sinha, Vibha Singhal and Saha, Diptikalyan and Dhoolia, Pankaj and Padhye, Rohan and Mani, Senthil},
    booktitle={2015 IEEE/ACM 12th Working Conference on Mining Software Repositories}, 
    title={Detecting and Mitigating Secret-Key Leaks in Source Code Repositories}, 
    year={2015},
    pages={396-400},
    doi={10.1109/MSR.2015.48}
}

@article{api-revoke,
  title={Advanced API Security Techniques and Service Management},
  volume={3},
  url={https://ijeret.org/index.php/ijeret/article/view/261},
  doi={10.63282/3050-922X.IJERET-V3I4P108}, 
  number={4},
  journal={International Journal of Emerging Research in Engineering and Technology},
  author={Jangam, Sandeep Kumar and Karri, Nagireddy and Pedda Muntala, Partha Sarathi Reddy},
  year={2022},
  month={Dec.},
  pages={63–74}
}

@InProceedings{ip-login-detect,
author="Saxon, James
and Feamster, Nick",
editor="Hohlfeld, Oliver
and Moura, Giovane
and Pelsser, Cristel",
title="GPS-Based Geolocation of Consumer IP Addresses",
booktitle="Passive and Active Measurement",
year="2022",
publisher="Springer International Publishing",
address="Cham",
pages="122--151",
isbn="978-3-030-98785-5",
doi={https://doi.org/10.1007/978-3-030-98785-5_6}
}

@INPROCEEDINGS{usage-pattern,
  author={Balasupramanian, N. and Ephrem, Ben George and Al-Barwani, Imad Salim},
  booktitle={2017 International Conference on Intelligent Computing, Instrumentation and Control Technologies (ICICICT)}, 
  title={User pattern based online fraud detection and prevention using big data analytics and self organizing maps}, 
  year={2017},
  volume={},
  number={},
  pages={691-694},
  doi={10.1109/ICICICT1.2017.8342647}}

@Inbook{sign-out-everywhere,
author="Wilson, Yvonne
and Hingnikar, Abhishek",
title="Logout",
bookTitle="Solving Identity Management in Modern Applications: Demystifying OAuth 2, OpenID Connect, and SAML 2",
year="2023",
publisher="Apress",
address="Berkeley, CA",
pages="219--231",
isbn="978-1-4842-8261-8",
doi="10.1007/978-1-4842-8261-8_13",
url="https://doi.org/10.1007/978-1-4842-8261-8_13"
}

@misc{x-api-leak,
    title = {{xAI Dev Leaks API Key for Private SpaceX, Tesla LLMs - Krebs on Security}},
    author = {Krebs on Security},
    howpublished = {\url{https://krebsonsecurity.com/2025/05/xai-dev-leaks-api-key-for-private-spacex-tesla-llms/}}
}

@misc{api-leak-cost,
    title = {{IT Leaders Share Cost of API Incidents, Concerns Over AI Threats | Kong Inc.}},
    author = {Kong Inc.},
    howpublished = {\url{https://konghq.com/blog/enterprise/cost-of-api-security-incidents-2025}}
}

@misc{ios-security,
      title = {{Modern iOS Security Features -- A Deep Dive into SPTM, TXM, and Exclaves}}, 
      author = {Moritz Steffin and Jiska Classen},
      year = {2025},
      eprint = {2510.09272},
      archivePrefix = {arXiv},
      primaryClass = {cs.CR},
      howpublished = {\url{https://arxiv.org/abs/2510.09272}}, 
}

@misc{ohttp,
    title = {{RFC 9458: Oblivious HTTP}},
    author = {Martin Thomson and Christopher A. Wood},
    howpublished = {\url{https://www.rfc-editor.org/rfc/rfc9458.html}}
}

@misc{privacy-pass-scheme,
    title = {{RFC 9577: The Privacy Pass HTTP Authentication Scheme}},
    author = {Tommy Pauly and Steven Valdez and Christopher A. Wood},
    howpublished = {\url{https://www.rfc-editor.org/rfc/rfc9577.html}}
}

@misc{privacy-pass,
    title = {{RFC 9578: Privacy Pass Issuance Protocols}},
    author = {Sofia Celi and Alex Davidson and Steven Valdez and Christopher A. Wood},
    howpublished = {\url{https://www.rfc-editor.org/rfc/rfc9578.html}}
}

@misc{amfi-exempt,
    title = {{osy/AMFIExemption: Grant private entitlements to OSX aps}},
    author = {osy},
    howpublished = {\url{https://github.com/osy/AMFIExemption}}
}

@misc{private-relay,
    title = {{iCloud Private Relay Overview}},
    author = {Apple},
    howpublished = {\url{https://www.apple.com/privacy/docs/iCloud_Private_Relay_Overview_Dec2021.PDF}},
    year = {2021}
}

@misc{keychain-pwd-1,
    title = {{What is my "login" keychain password? My ... - Apple Community}},
    author = {Apple Community},
    howpublished = {\url{https://discussions.apple.com/thread/254424961?sortBy=rank}}
}

@misc{keychain-pwd-2,
    title = {{Microsoft Outlook for Mac - Apple Community}},
    author = {Apple Community},
    howpublished = {\url{https://discussions.apple.com/thread/252714597?sortBy=rank}}
}

@misc{keychain-pwd-3,
    title = {{This keeps popping up how do i get rid of this: r/mac}},
    author = {Reddit},
    howpublished = {\url{https://www.reddit.com/r/mac/comments/zuqdzi/this_keeps_popping_up_how_do_i_get_rid_of_this/}}
}

@misc{CVE-2017-7150,
    title = {{CVE Record: CVE-2017-7150}},
    author = {Patrick Wardle},
    howpublished = {\url{https://www.cve.org/CVERecord?id=CVE-2017-7150}}
}

@misc{amos,
    title = {{Atomic Stealer: Dissecting 2024's Most Notorious macOS Infostealer}},
    author = {Sıla Özeren Hacıoğlu},
    howpublished = {\url{https://www.picussecurity.com/resource/blog/atomic-stealer-amos-macos-threat-analysis}}
}

@misc{keychain,
    author  = {Apple},
    title   = {{Keychain Services}},
    year    = {2025},
    howpublished = {\url{https://developer.apple.com/documentation/security/keychain-services}}
}

@misc{restricted-entitlements,
    author  = {Apple},
    title   = {{Creating distribution-signed code for macOS | Apple Developer Documentation}},
    howpublished = {\url{https://developer.apple.com/documentation/xcode/creating-distribution-signed-code-for-the-mac}}
}

@misc{sep,
    author  = {Apple},
    title   = {{Secure Enclave - Apple Support}},
    year    = {2024},
    howpublished = {\url{https://support.apple.com/guide/security/secure-enclave-sec59b0b31ff/web}}
}

@inproceedings{mini-apps,
  author    = {Zhang, Jiale and Zhang, Yue and Yang, Yuqing and Lin, Zhiqiang},
  title     = {{The Skeleton Keys: A Large Scale Analysis of Credential Leakage in Mini-apps}},
  booktitle = {Proceedings of the Network and Distributed System Security Symposium (NDSS)},
  year      = {2025},
  publisher = {The Internet Society},
  address   = {San Diego, CA}
}

@article{android-leak,
author = {Wei, Lili and Huang, Heqing and Cheung, Shing-Chi and Li, Kevin},
title = {How far are app secrets from being stolen? a case study on android},
year = {2025},
issue_date = {Mar 2025},
publisher = {Kluwer Academic Publishers},
address = {USA},
volume = {30},
number = {3},
issn = {1382-3256},
url = {https://doi.org/10.1007/s10664-024-10607-9},
doi = {10.1007/s10664-024-10607-9},
journal = {Empirical Softw. Engg.},
month = apr,
numpages = {32}
}

@INPROCEEDINGS{ios-leak,
  author={Wen, Haohuang and Li, Juanru and Zhang, Yuanyuan and Gu, Dawu},
  booktitle={2018 25th Asia-Pacific Software Engineering Conference (APSEC)}, 
  title={An Empirical Study of SDK Credential Misuse in iOS Apps}, 
  year={2018},
  volume={},
  number={},
  pages={258-267},
  doi={10.1109/APSEC.2018.00040}
}

@article{privacy-pass-paper,
    title = "Privacy Pass: Bypassing Internet Challenges Anonymously",
    author = "Alexander Davidson",
    year = "2018",
    doi = "10.1515/popets-2018-0026",
    language = "English",
    volume = "2018",
    pages = "164--180",
    journal = "Proceedings on Privacy Enhancing Technologies",
    issn = "2299-0984",
    publisher = "de Gruyter",
    number = "3",
}

@misc{privacy-pass-verification,
      author = {Kristiana Ivanova and Daniel Gardham and Stephan Wesemeyer},
      title = {Formal Verification of Privacy Pass},
      howpublished = {Cryptology {ePrint} Archive, Paper 2025/2022},
      year = {2025},
      url = {https://eprint.iacr.org/2025/2022}
}

@INPROCEEDINGS{OPRF,
    author={Casacuberta, Sílvia and Hesse, Julia and Lehmann, Anja},
    booktitle={2022 IEEE 7th European Symposium on Security and Privacy (EuroS\&P)}, 
    title={SoK: Oblivious Pseudorandom Functions}, 
    year={2022},
    pages={625-646},
    doi={10.1109/EuroSP53844.2022.00045}
}

@inproceedings{sinha,
  author    = {Sinha, Vibha Singhal and Saha, Diptikalyan and Dhoolia, Pankaj and Padhye, Rohan and Mani, Senthil},
  title     = {{Detecting and Mitigating Secret-Key Leaks in Source Code Repositories}},
  booktitle = {Proceedings of the 12th Working Conference on Mining Software Repositories (MSR)},
  year      = {2015},
  pages     = {396--400},
  publisher = {IEEE},
  doi       = {10.1109/MSR.2015.48}
}

@inproceedings{secrethunter,
  author    = {Wen, Elliott and Wang, Jia and Dietrich, Jens},
  title     = {{SecretHunter: A Large-scale Secret Scanner for Public Git Repositories}},
  booktitle = {Proceedings of the 2022 IEEE International Conference on Trust, Security and Privacy in Computing and Communications (TrustCom)},
  year      = {2022},
  pages     = {123--130},
  publisher = {IEEE},
  doi       = {10.1109/TrustCom56396.2022.00028}
}

@inproceedings{lounici,
  author    = {Lounici, Sofiane and Rosa, Marco and Negri, Carlo Maria and Trabelsi, Slim and \"{O}nen, Melek},
  title     = {{Optimizing Leak Detection in Open-source Platforms with Machine Learning Techniques}},
  booktitle = {Proceedings of the 7th International Conference on Information Systems Security and Privacy (ICISSP)},
  year      = {2021},
  pages     = {145--159},
  publisher = {SCITEPRESS},
  doi       = {10.5220/0010238101450159}
}

@inproceedings{saha,
  author    = {Saha, Aakanksha and Denning, Tamara and Srikumar, Vivek and Kasera, Sneha Kumar},
  title     = {{Secrets in Source Code: Reducing False Positives Using Machine Learning}},
  booktitle = {Proceedings of the 2020 International Conference on COMmunication Systems \& NETworkS (COMSNETS)},
  year      = {2020},
  pages     = {168--175},
  publisher = {IEEE},
  doi       = {10.1109/COMSNETS48256.2020.9027350}
}

@ARTICLE{cookie-theft,
  author={Kwon, Hyunsoo and Nam, Hyunjae and Lee, Sangtae and Hahn, Changhee and Hur, Junbeom},
  journal={IEEE Transactions on Information Forensics and Security}, 
  title={(In-)Security of Cookies in HTTPS: Cookie Theft by Removing Cookie Flags}, 
  year={2020},
  volume={15},
  number={},
  pages={1204-1215},
  doi={10.1109/TIFS.2019.2938416}}

@INPROCEEDINGS{sso-theft,
  author={Wang, Rui and Chen, Shuo and Wang, XiaoFeng},
  booktitle={2012 IEEE Symposium on Security and Privacy}, 
  title={Signing Me onto Your Accounts through Facebook and Google: A Traffic-Guided Security Study of Commercially Deployed Single-Sign-On Web Services}, 
  year={2012},
  volume={},
  number={},
  pages={365-379},
  doi={10.1109/SP.2012.30}}

@inproceedings{cookie-pharming,
author = {Karlof, Chris and Shankar, Umesh and Tygar, J. D. and Wagner, David},
title = {Dynamic pharming attacks and locked same-origin policies for web browsers},
year = {2007},
isbn = {9781595937032},
publisher = {Association for Computing Machinery},
address = {New York, NY, USA},
url = {https://doi.org/10.1145/1315245.1315254},
doi = {10.1145/1315245.1315254},
booktitle = {Proceedings of the 14th ACM Conference on Computer and Communications Security},
pages = {58–71},
numpages = {14},
location = {Alexandria, Virginia, USA},
series = {CCS '07}
}

@inproceedings{csrf,
author = {Barth, Adam and Jackson, Collin and Mitchell, John C.},
title = {Robust defenses for cross-site request forgery},
year = {2008},
isbn = {9781595938107},
publisher = {Association for Computing Machinery},
address = {New York, NY, USA},
url = {https://doi.org/10.1145/1455770.1455782},
doi = {10.1145/1455770.1455782},
booktitle = {Proceedings of the 15th ACM Conference on Computer and Communications Security},
pages = {75–88},
numpages = {14},
location = {Alexandria, Virginia, USA},
series = {CCS '08}
}

@inproceedings{csrf-2,
author = {Shernan, Ethan and Carter, Henry and Tian, Dave and Traynor, Patrick and Butler, Kevin},
title = {More Guidelines Than Rules: CSRF Vulnerabilities from Noncompliant OAuth 2.0 Implementations},
year = {2015},
isbn = {9783319205496},
publisher = {Springer-Verlag},
address = {Berlin, Heidelberg},
url = {https://doi.org/10.1007/978-3-319-20550-2_13},
doi = {10.1007/978-3-319-20550-2_13},
booktitle = {Proceedings of the 12th International Conference on Detection of Intrusions and Malware, and Vulnerability Assessment - Volume 9148},
pages = {239–260},
numpages = {22},
location = {Milan, Italy},
series = {DIMVA 2015}
}

@inproceedings{session-hijacking,
author = {Johns, Martin and Braun, Bastian and Schrank, Michael and Posegga, Joachim},
title = {Reliable protection against session fixation attacks},
year = {2011},
isbn = {9781450301138},
publisher = {Association for Computing Machinery},
address = {New York, NY, USA},
url = {https://doi.org/10.1145/1982185.1982511},
doi = {10.1145/1982185.1982511},
booktitle = {Proceedings of the 2011 ACM Symposium on Applied Computing},
pages = {1531–1537},
numpages = {7},
location = {TaiChung, Taiwan},
series = {SAC '11}
}

\end{document}